# Switching dynamics and improved efficiency of free-standing antiferroelectric capacitors

*Umair Saeed[1,2*], David Pesquera[1*], Ying Liu[1,3], Ignasi Fina[4], Saptam Ganguly[1], José Santiso[1], Jessica Padilla[1], José Manuel Caicedo Roque[1] , Xiaozhou Liao[3], Gustau Catalan[1,5]*

[1]Catalan Institute of Nanoscience and Nanotechnology (ICN2), CSIC and BIST, Campus UAB, Bellaterra, Barcelona, 08193 Catalonia.

[2] Universitat Autònoma de Barcelona, Plaça Cívica, 08193 Bellaterra, Barcelona, Catalonia.

[3] School of Aerospace, Mechanical & Mechatronic Engineering, The University of Sydney, Sydney, NSW, 2006, Australia.

[4]Institute of Materials Science of Barcelona (ICMAB-CSIC), Campus UAB, Bellaterra, Barcelona, 08193 Catalonia.

[5]ICREA - Institució Catalana de Recerca i Estudis Avançats, Barcelona, Catalonia, 08010.

E-mail:      umair.saeed@icn2.cat
             gustau.catalan@icn2.cat

We report the switching dynamics of antiferroelectric Lead Zirconate ($PbZrO_3$) free standing capacitors compared to their epitaxial counterparts. Frequency dependence of hysteresis indicates that freestanding capacitors exhibit a lower dispersion of switching fields, lower residual polarization, and faster switching response as compared to epitaxially-clamped capacitors. As a consequence, freestanding capacitor membranes exhibit better energy storage density and efficiency.

# 1. Introduction

Antiferroelectric (AFE) materials have gained traction in the scientific community owing to actual or potential applications in high energy-storage density capacitors,[1] solid-state cooling via a negative electrocaloric effect,[2] micro-electro-mechanical systems (MEMS),[3] and photovoltaics.[4] These applications are based on field-induced dipole ordering. Unlike ferroelectric materials, the dipoles in AFEs are arranged antiparallel within the unit cell, and therefore do not show spontaneous polarization; however, under a strong enough external electric field, the dipoles reorient themselves and undergo an AFE to Ferroelectric (FE) phase transition, resulting in a sharp increase of polarization, strain, and latent heat.[5] This phase transition is reversible in most AFEs, giving rise to a double hysteresis loop of polarization vs electric field. Lead Zirconate- $PbZrO_3$ (PZO) was the first antiferroelectric to be discovered already more than 70 years ago,[6] and it has since been considered the archetype AFE.

The field-induced AFE-FE transition of PZO involves a change in crystal class (from orthorhombic to rhombohedral),[7] which causes ferroelastic twinning in addition to changes to the unit cell volume. Accordingly, the transition is sensitive to mechanical stress and hence to epitaxial strain. Epitaxially-grown thin films allow controlling orientation and strain by using different substrates[8] and, in AFEs, control over these parameters can lead to higher energy storage density than ceramics.[9] On the other hand, epitaxial strain comes with a cost: it changes the free-energy balance in favor of the ferroelectric phase, and for thin strained films, PZO can be FE instead of AFE.[10] At intermediate regimes of strain and/or external field, ferrielectricity (FiE) has been reported,[11] further complicating the phase diagram. Moreover, since the AFE-FE transition involves a change in spontaneous strain, clamping to a rigid substrate can affect switching by hindering the development of the spontaneous strain concomitant with the transition. Hence, while the ability of epitaxy to produce well-oriented, high quality thin films is desirable, removing substrate clamping may also be beneficial.

In this context, the development of water-soluble sacrificial layers that are epitaxially coherent with perovskite substrates[12] opens a new avenue for research and applications, by retaining the orientation control of the growth while allowing to remove the substrate clamping –in addition to facilitating integration onto semiconductors and flexible electronics. This strategy has proven fruitful on ferroelectric oxides, such as Barium Titanate[12b,13] and Bismuth Ferrite,[14] where substrate removal allowed demonstrating large flexibility in the freestanding membranes and improved performance (reduced switching energy and increased switching speeds) in integrated devices. In contrast, research on free-standing antiferroelectric films has only just started,[10a,15] and the effect of substrate removal on the switching properties of antiferroelectric free-standing capacitors is as yet unknown. In addition, the only free-standing capacitors made so far [REF] had ex-situ deposited base-metal electrodes, rather than the heteroepitaxial oxide electrodes normally used in capacitor thin films, which complicates separating the effects of strain and clamping from the effect of electrodes. Making free-standing antiferroelectric capacitors with coherent electrodes and determining how their properties differ from epitaxially clamped ones is the purpose of the present paper.

In this work, we have fabricated heteroepitaxial free-standing capacitors of antiferroelectric $PbZrO_3$ (PZO) sandwiched between $SrRuO_3$ (SRO) electrodes. These were grown on (110)-oriented $GdScO_3$ (GSO) single crystal substrates, with a sacrificial water-soluble layer of $Sr_3Al_2O_6$ (SAO) that was then dissolved. Nominally identical capacitor structures were also grown directly on the perovskite substrates, thus allowing comparison between free-standing and clamped capacitors. We then characterized the structure, electrical properties and switching dynamics on both systems. We find that removal of the mechanical clamping of the substrate has a beneficial impact on the switching dynamics (which becomes faster) and energy storage performance (which becomes higher).

## 2. Results and discussions

### 2.1. Synthesis

Tri-layer epitaxial heterostructures consisting of 300 nm PZO thin films with top and bottom 35 nm SRO electrodes were grown on (110)-oriented (GSO) single crystal substrates via pulsed laser deposition (PLD) (growth details provided in Experimental Section). Tri-layers that were deposited directly on the GSO substrate are from now on referred to as "on GSO". We also grew similar capacitor structures on SAO-buffered substrates. The SAO layer was subsequently etched in water, releasing the capacitor structure that was transferred to a metal-coated silicon substrate, schematic shown in **Figure 1b** (details in **Experimental Section**). These membranes adhere to the silicon supports by virtue of van der Waals forces. The membrane capacitors released from GSO are henceforth referred to as ¨off GSO".

To perform electrical measurements in the on-GSO epitaxial capacitors, the top SRO electrodes were patterned as circular discs (10-400 µm diameter) (**Figure 1a and 1c**). For the off-GSO membrane capacitors, electrode patterning was not required, because during the transfer the membranes crack along the PZO crystallographic directions, readily providing rectangular-shaped small capacitors with dimensions varying between 10 and 100 um lateral size (**Figure 1d**). After synthesis, all samples were annealed at 260°C for 15 minutes, with a heating and cooling rate of 2°C min$^{-1}$.

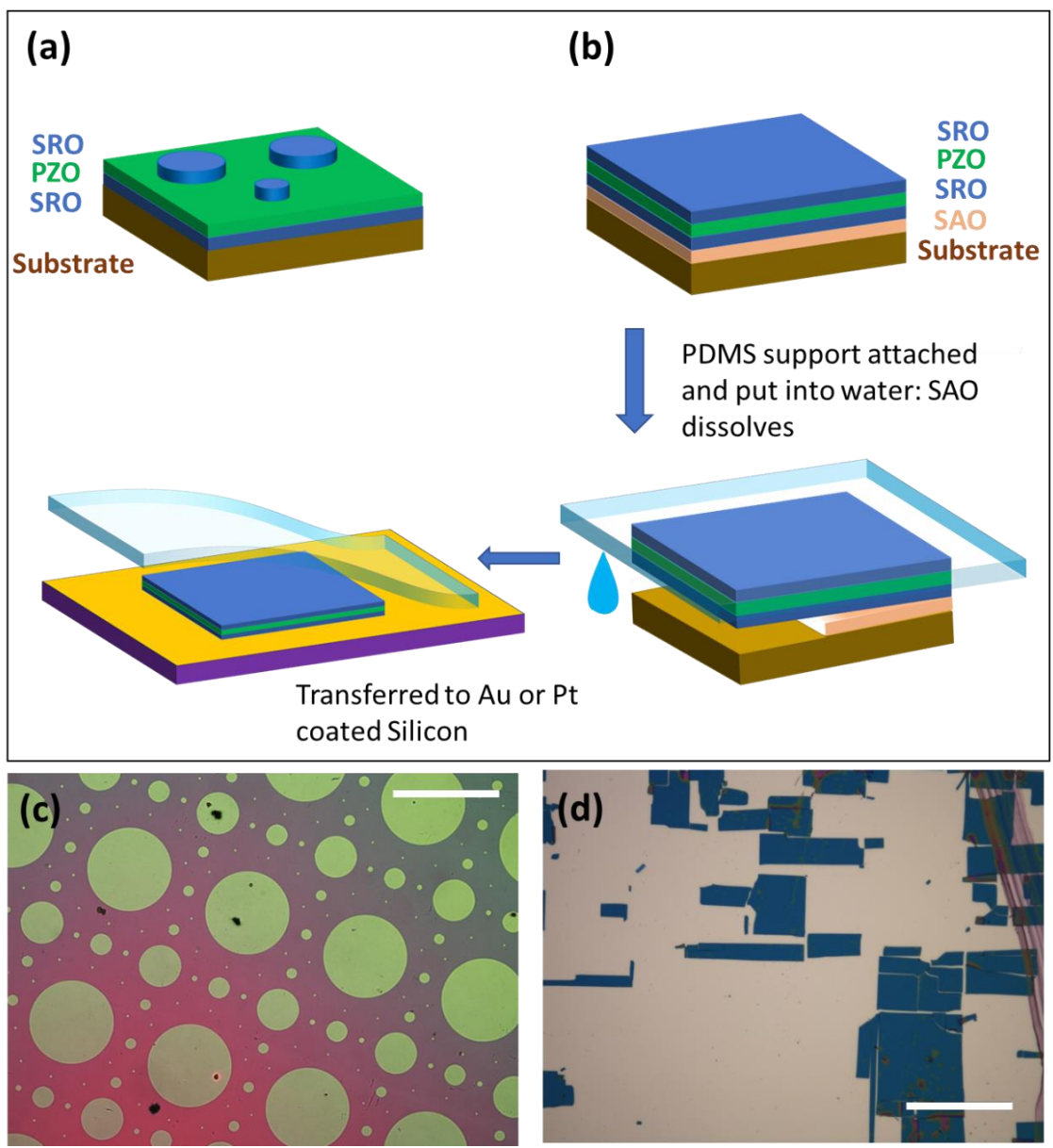

**Figure 1: (a)**: Schematic of patterned SRO top electrodes on epitaxial heterostructure, **(b):** Fabrication of SRO/PZO/SRO free-standing membranes by PDMS release method. **(c):** Patterned SRO top electrodes on epitaxial heterostructure under a light microscope (scale bar: 500 µm), **(d):** Free-standing capacitors on Au coated Silicon substrate under a light microscope (scale bar: 100 µm).

## 2.2. Structure

We examined the "Off GSO" membrane (**Figure 2**) under cross-sectional scanning transmission electron microscopy (STEM) along $[001]_{PC}$ zone axis. The images show a PZO thickness of 300 nm and reveal that the PZO layer grew via columnar growth mode, which causes dislocations to form along the boundaries of the columns. Mappings for lead atom displacement with respect to surrounding Zr atoms **(Figures 2 (b))** and geometric phase analysis (GPA) **(Figures 2(c))** were calculated from high angle annular dark field (HAADF) images taken in selected areas with [120]$_O$ orientation, where the anti-polar direction is oriented 45 degrees out of the surface plane. The antiparallel dipole arrangement of two up – two down typical of the antiferroelectric Pbam phase is observed in most of the analyzed area. In addition to the AFE phase, ferrielectric modulations of the dipole magnitudes (red circled areas in **Figure 2 (b)**) and translational boundaries (dashed white lines) with uncompensated dipole sublattices were observed. Overall, the atomic structure of the membranes is similar to what has been observed in bulk crystals and ceramics[16] . The TEM results were virtually the same for the epitaxial films (**Supplementary information S1**). The TEM analysis therefore indicates that both the epitaxially clamped and the released capacitors have the same room-temperature antiferroelectric structure as bulk, with minor ferrielectric inclusions.

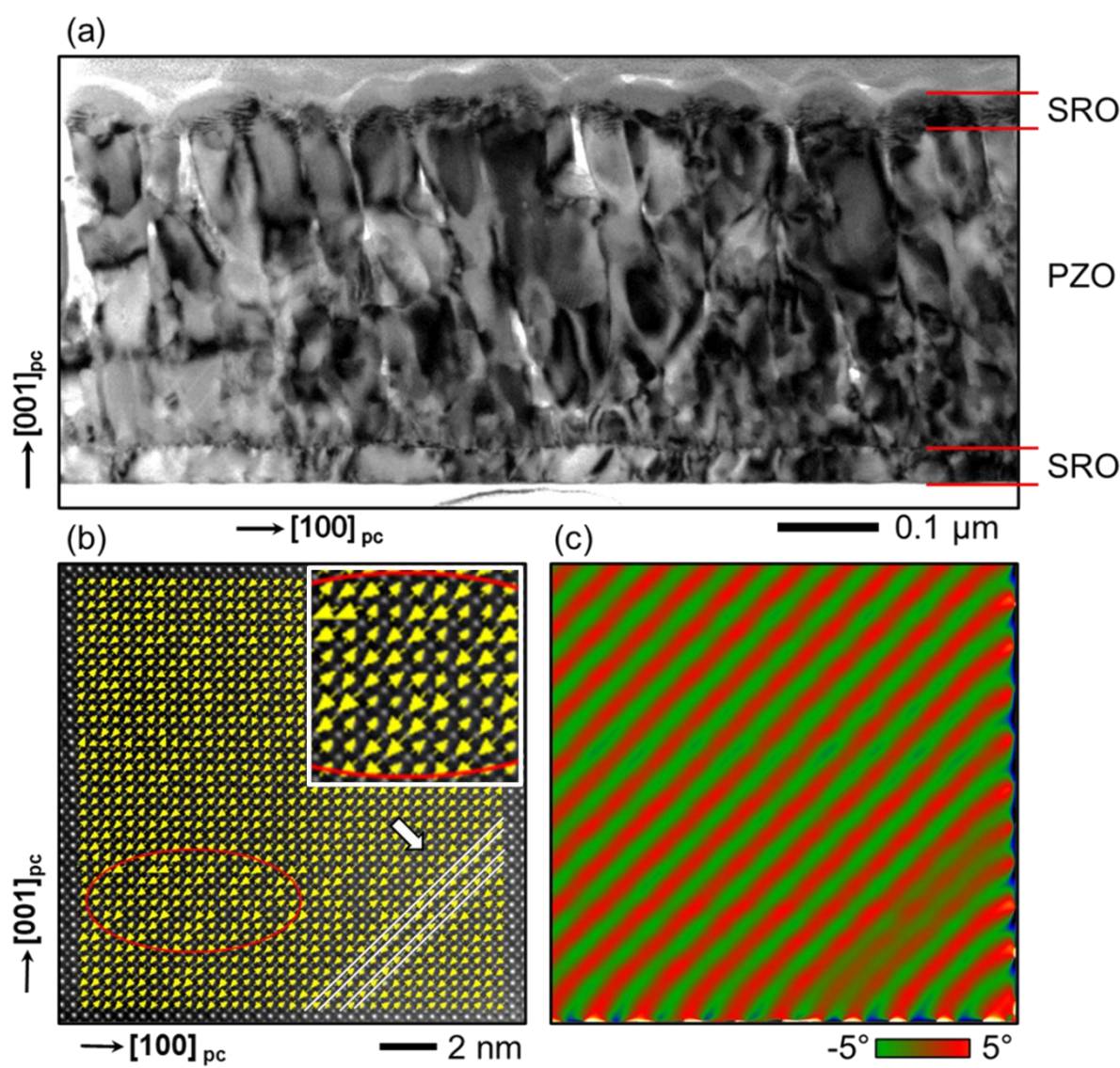

**Figure 2**: STEM of "Off GSO" membrane, where **(a)** Low resolution TEM image, **(b)** Lead atom displacement mapping superimposed on HAADF TEM image and inset showing zoom in of the red circle, **(c)** GPA rotation maps. The area enclosed in red circle shows a FiE region with unequal opposite Pb atom displacement, white arrow represents a translational boundary.

To gain quantitative insight into the structure and strain state of the capacitors, we use High-Resolution X-ray Diffraction (HRXRD) (**Figure 3a**, Experimental Section). Both types of samples show a coherent crystallographic orientation with the perovskite pseudocubic $(100)_{pc}$ peak out-of-plane, in agreement with the TEM results. At room temperature, the symmetry of PZO is orthorhombic, which results in a splitting of the (100)pc peak into two domains: $(001)_O$ and $(120)_O$, where $(001)_O$- oriented domains have the antipolar axis in-plane, while $(120)_O$ domains have the dipole orientation along the $[110]/[-110]_{pc}$ axis, and therefore an out-of-plane antipolar component. **Figure 3(c)** sketches the two types of orientations that are present with respect to the surface plane. The relative intensities of the X-rays allow quantifying the relative population of the different antiferroelectric domains. On-GSO epitaxial capacitors show higher proportion of $(001)_O$ orientation than $(120)_O$, and incorporating SAO within this heterostructure does not seem to affect the orientation proportions before release from the substrate (compare red and black diffractogram in **figure 3a**); however, after release, the membranes show a higher proportion of $(120)_O$ (blue diffractogram in **figure 3a**).

The volume fraction of each type of domain was obtained from the integrated area of each diffraction peak and the average strains were obtained from the diffraction peak positions. Using Bragg´s law, the average out-of-plane pseudocubic lattice parameters are determined, and the percentage strain for each domain orientation is calculated as:

$$\varepsilon_{(hkl)} = \frac{a_{(hkl)_{sample}} - a_{(hkl)_{bulk}}}{a_{(hkl)_{bulk}}} * 100 \qquad (1)$$

In addition, inhomogeneous strain resulting from local disorder was obtained from the peak widths via Williamson-Hall plots (**Figure 3b**), whereby the peak width is plotted as a function of peak angle and fitted to the relation[17]:

$$(\beta_{sample} \cos(\theta))^2 = (\frac{K\lambda}{D})^2 + (4\varepsilon_i \sin(\theta))^2 \qquad (2)$$

Where the first term on the right is related to size effects (*D* is the coherence length in the out-of-plane direction, and the smaller this coherent length the bigger the broadening), while the second term is related to the inhomogeneous strain $\varepsilon_i$, $\lambda$ is the XRD wavelength, *K* is a geometric constant depending on the shape of the crystal and equals 0.94 for spherical crystals with cubic symmetry.[17] All this information is summarized in **Table 1**.

The results indicate that, while some residual strain is relaxed upon release from the substrate, the epitaxial film is already quite relaxed before release (the strain in the PZO domains of the epitaxial film is between -0.2% and -0.3%, whereas the theoretical lattice mismatch with the substrate is more than 4%. Since the epitaxial films are already strain-relaxed, the average strain difference between epitaxial film and released membrane is marginal (0.089%). Likewise, the inhomogeneous strain is only marginally smaller in the released capacitor compared to the epitaxial one (0.13 (± 0.005) % for "On-GSO" film vs 0.11 (± 0.002) % for "Off-GSO" membrane). We therefore surmise that release from the substrate has only a small effect on the strain state, in spite of which there is a large difference in functional properties, as we will see next. The main effect of the substrate is therefore not so much to strain the film, but to clamp it, with the interfacial pinning affecting the switching dynamics.

|  | Peak | Volume fraction | Lattice parameter (Å) | Bulk Lattice Parameter (Å) [18] | Strain (%) | Inhomogeneous strain (%) |
|---|---|---|---|---|---|---|
| **On GSO** | $(002)_{PC} = (004)_O$ | 0.79 | 4.118 | 4.126 | -0.199 | 0.13 |
|  | $(200)_{PC} = (240)_O$ | 0.21 | 4.156 | 4.168 | -0.288 | 0.12 |
| **Off GSO** | $(002)_{PC} = (004)_O$ | 0.29 | 4.122 | 4.126 | -0.102 | 0.11 |
|  | $(200)_{PC} = (240)_O$ | 0.71 | 4.162 | 4.168 | -0.144 | 0.11 |

**Table 1:** Calculated Lattice parameters and strains for film and membrane capacitors.

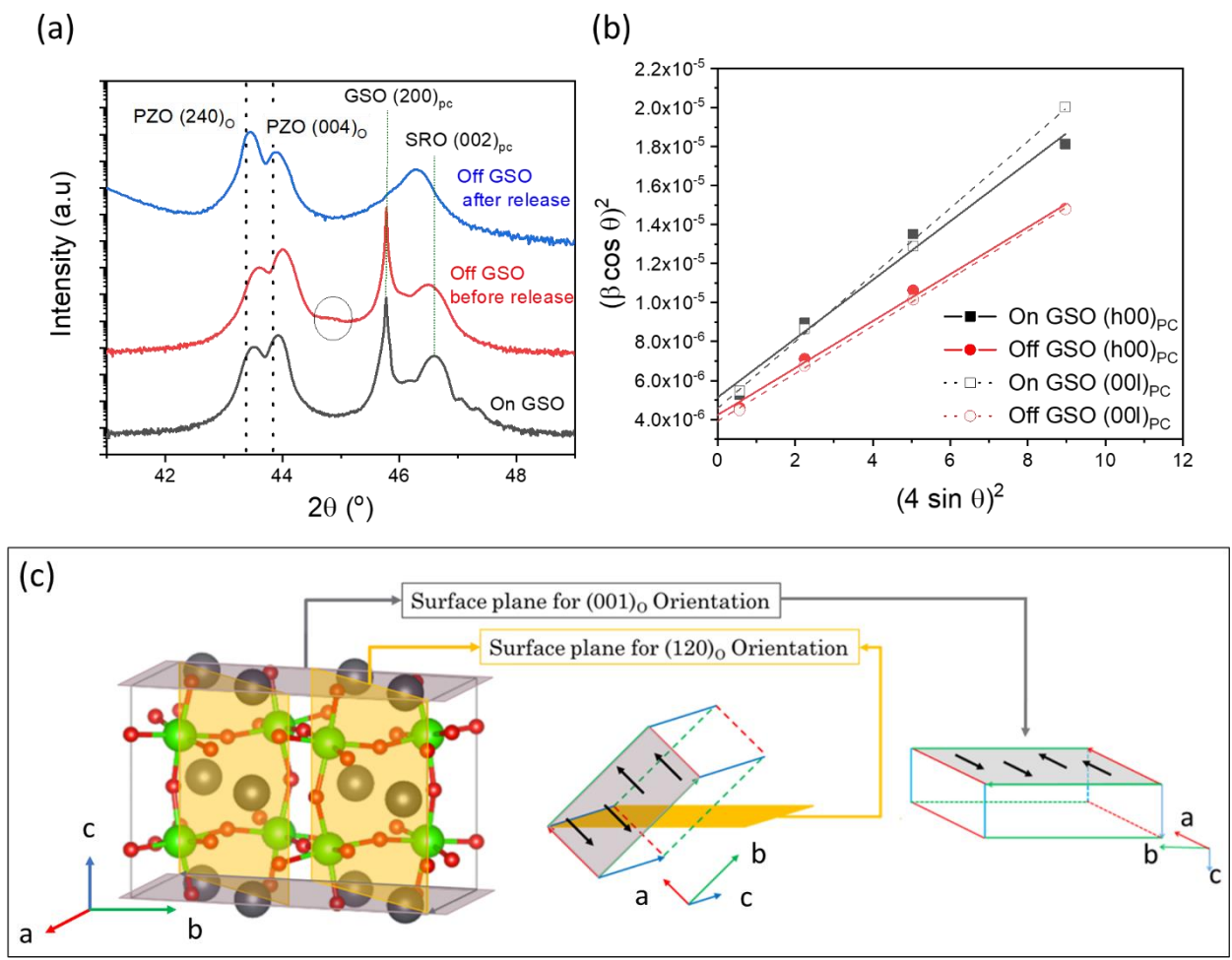

**Figure 3**: HRXRD of heterostructures grown on **(a):** GSO substrates, black lines for "On GSO" films, red for heterostructure with SAO before release, and blue for "Off GSO" membranes after release, the black dotted lines on the PZO peaks represent the bulk 2-theta value for each orientation, circle in the red curve shows the SAO peak, **(b)**: Williamson-Hall type plot for "On GSO" films and "Off GSO" membranes, **(c):** Representation of $(120)_O$ and $(001)_O$ orientations, black circles represent Pb atoms, green are Zr atoms, and red are O atoms, black arrows represent orientation of dipole moments.

## 2.3. Switching dynamics

Double-hysteresis loops for polarization as a function of voltage were measured for both types of capacitors at 10 kHz at room temperature and are shown in **Figure 4.** There are notable differences between the loops of epitaxial films and free-standing membrane capacitors. Compared to the membrane, the "On GSO" film shows a lower switching field, slightly higher saturation polarization, and a considerable remnant polarization despite being structurally antiferroelectric, as evidenced from STEM as well as the four characteristic current peaks. By contrast, the membrane capacitor shows a much clearer double hysteresis, closer to the ideal antiferroelectric loop.

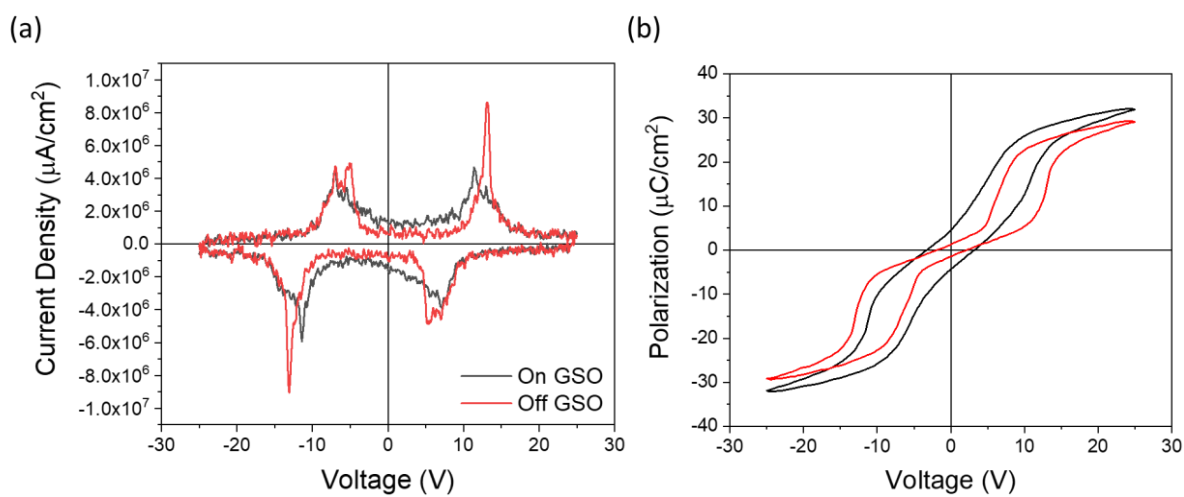

**Figure 4**: Comparison of **(a)**: I-V curves and **(b):** Polarization loops of epitaxial film "on GSO" vs membrane "off GSO" at 10 kHz.

Since the XRD and TEM structural investigation does not show a substantial difference in crystal symmetry or in concentration of ferrielectric translational boundaries, we therefore do not think that the larger "remnant polarization" of the epitaxial film is indicative of a stable polar phase. The polarization at 0V is mostly metastable, reflecting a difference in switching dynamics whereby epitaxial films take longer to relax back to the AFE state. To substantiate this claim, we examine the switching hysteresis loops as a function of loop frequency.

The current density vs electric field curves are shown in **Figure 5 (a- b)**, and the corresponding polarization loops for some frequencies are shown in the **Supplementary Material S3**. The AFE-FE switching peaks in the membranes are sharper compared to the epitaxial films, indicating lower disorder (lower spread of activation energies).[19] Interestingly, the FE-AFE return peaks display two distinct switching events for the membranes, suggesting the existence of an intermediate stage or bridging phase during switching.[20] In the epitaxial films, the switching peaks are so broad that the intermediate step, if present, is subsumed and cannot be singled out.

The positive and negative AFE-FE and FE-AFE switching fields as a function of hysteresis loop frequency were extracted from the current-voltage curves and fitted with equations based on the Kolmogorov-Avrami-Ishibashi (KAI) model: [21]

$$E_{AFE-FE} = A + K_A * f^\beta \qquad (3)$$
$$E_{FE-AFE} = F - K_F * f^\beta \qquad (4)$$

where *A* and *F* are the quasistatic electric fields for AFE to FE and FE to AFE switching, which are related to the energy barrier for the transition, and $K_A$, $K_F$ and $\beta$ are dynamic parameters related to the mobility of domain walls or phase boundaries. We ruled out contact resistance differences (**Supplementary material S4**), which is important because the apparent exponents in these fittings can otherwise be influenced by electronic time constants (RC) of the capacitors. As mentioned, the FE-AFE switching of the membranes has two steps. We focused on the most intense of the two switching peaks; the smaller one was also fitted and showed a similar trend with a slightly slower relaxation time (**Supplementary material S6**), so the conclusions of the present analysis apply equally to both. The results of the fitting are summarized in Table 2.

| **Fitting Parameters** | | **On GSO** | **Off GSO** |
|---|---|---|---|
| $E_{AFE-FE} = A + K_A * f^\beta$ | A | 330 | 404 |
| | $K_A$ | 3.80 | 5.17 |
| | β | 0.88 | 0.80 |
| $E_{FE-AFE} = F - K_F * f^\beta$ | F | 253 | 222 |
| | $K_F$ | 2.90 | 10.3 |
| | β | 0.96 | 0.62 |

**Table 2:** Values of fitted parameters from **Equation 3 and 4**.

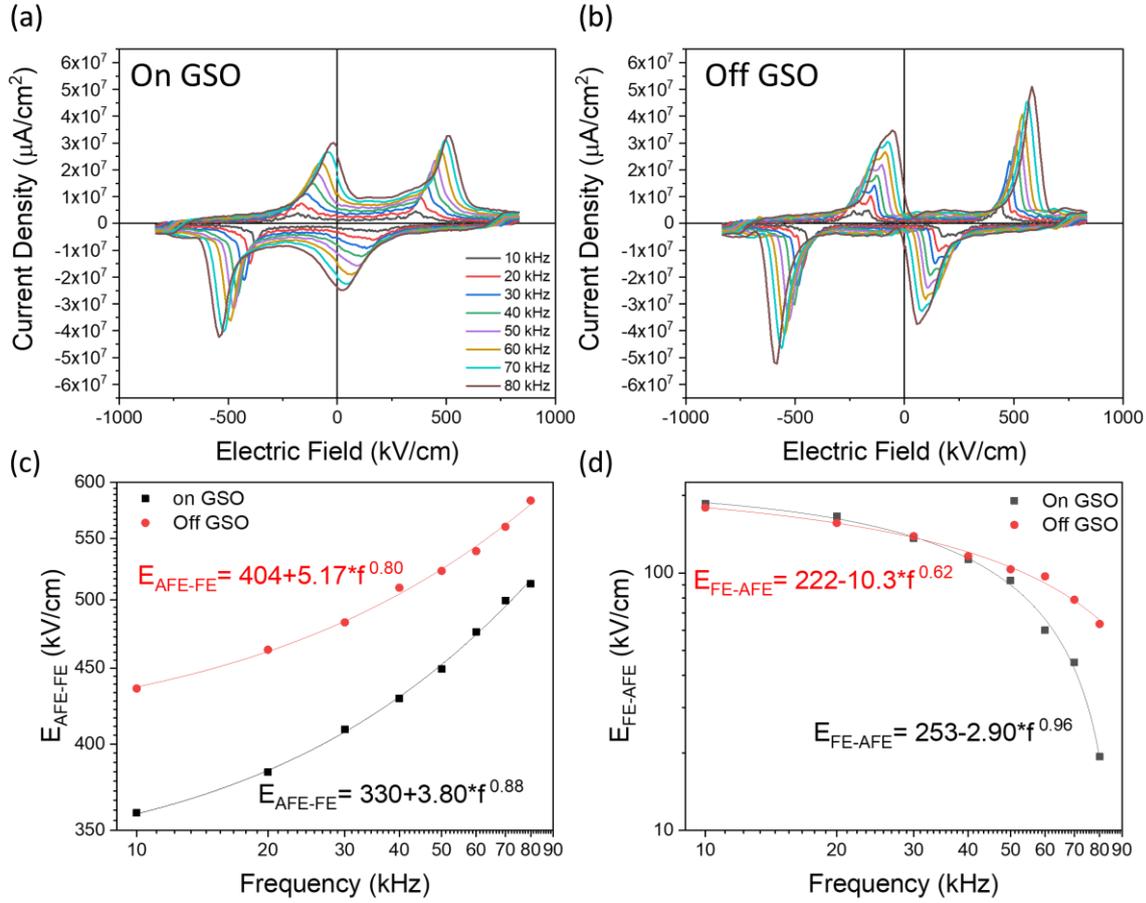

**Figure 5**: Current density – Electric field curves of, **(a):** "on GSO" film, **(b):** "off GSO" membrane, and KAI model fittings for **(c)**: AFE- FE switching fields and, **(d)**: FE- AFE switching fields.

The parameters *A* and *F* represent the critical fields in the static limit. We observe a higher value of A in membranes compared to epitaxial films, suggesting that the energy barrier for the AFE-FE transition is higher in the membranes or, more likely, that the AFE-FE energy barrier is lowered by disorder or interfacial nucleation defects such as strain hotspots in the epitaxial film. In contrast, the critical static field for the reverse FE-AFE transition (F) is very similar for both capacitors.

The frequency dependence is captured by the coefficient K and, especially, by the dynamic exponent *β*, whose impact on critical field increases exponentially with increasing frequency. Here, we observe that, whereas in field for the FE-AFE transition is similar for films and membranes, at high frequencies the difference is dramatic. Due to the higher value of *β* in the epitaxial film (*β* ~1 compared to *β=0.6* for the membrane), at high frequencies the FE-AFE field of the epitaxial film drops very rapidly, becoming 0 at a critical frequency $f_c$ that can be calculated by setting $E_{FE\text{-}AFE} = 0$ in **Equation 4**, which leads to

$$f_c = \left(\frac{F}{K_F}\right)^{1/\beta} \tag{5}$$

For the epitaxial capacitor, $f_c$ is 90 kHz; at any frequency higher than that, most of the polarization will remain dynamically trapped in the ferroelectric state at 0V. By contrast, the membranes have a more agile antiferroelectric response with a higher cutoff frequency of 144 kHz. The difference is due to the higher value of $\beta$ in the epitaxial films, consistent with higher domain-wall-pinning disorder,[18] which causes the transition to become more viscous. Conversely, the phase boundaries in the membranes experience less friction (less interfacial pinning), allowing the hysteresis to respond more fluently.

Another interesting characteristic frequency can be defined if the rate at which $E_{FE\text{-}AFE}$ decreases is faster than the rate at which $E_{AFE\text{-}FE}$ increases. If that happens, there is a finite frequency, $f_F$, at which the positive FE-AFE field (at which the polarization goes $+P_{ferro}$ to 0) coincides with the negative AFE-FE field (at which polarization goes from 0 to $-P_{ferro}$), and therefore the PZO capacitor will switch directly between upwards and downwards polarized ferroelectric states without transitioning through the antiferroelectric state at all. To calculate $f_F$ we set :

$$E_{FE-AFE}(f_F) = -E_{AFE-FE}(f_F)$$

For the "On GSO" film, this frequency is $f_F$ ~1 MHz, and the "ferroelectric-like" coercive field ($E_{FE\text{-}AFE}(f_F)$) is 2.5 MV/cm. In contrast, the "off GSO" membrane does not exhibit this behavior because the frequency dispersion of $E_{AFE\text{-}FE}$ is higher than that of $E_{FE\text{-}AFE}$, so the second field never catches the first.

### 2.4 Dielectric response

The structural and functional analysis indicates that (i) there is not much difference in strain between epitaxial films and free-standing ones, in spite of which (ii) the epitaxial films have a higher "viscosity" in their switching dynamics, with a bigger frequency dispersion and slower response than the membranes. These observations are reflected by the behavior of the dielectric constant as a function of temperature (**Figure 6 (a)**) and amplitude of the AC electric field (**Figure 6 (b)**). Both capacitor morphologies show a transition at the Curie temperature of 510 K, which is very close to the bulk value,[22] consistent with the fact that both capacitors are strain-free. It is also worth noticing that the epitaxial film has a broader peak, consistent with bigger disorder. The dielectric losses are low for both capacitors too, confirming their high quality and low leakage, but the loss is lower in the membrane, consistent with the lower domain wall pinning anticipated from the hysteresis comparisons.

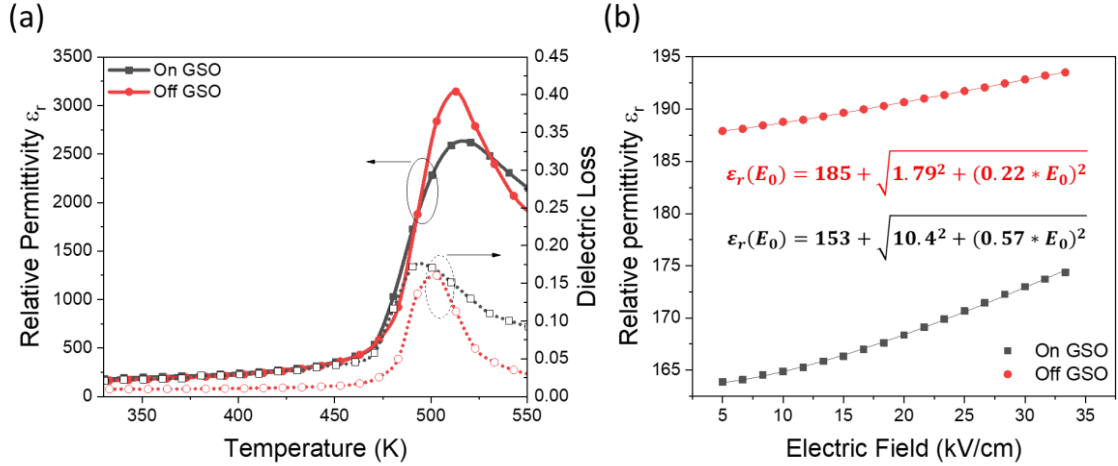

**Figure 6: (a)** Relative permittivity and dielectric loss measurements, "On GSO" film and "Off GSO" membrane, **(b)** Fitting of relative permittivity as a function of applied electric field.

To verify and quantify the higher viscosity of the transition in clamped films compared to released membranes, we have calculated the Rayleigh´s coefficients[23]. The relative permittivity is measured at different alternating fields (lower than the critical field) and fit to a hyperbolic law as shown in **Equation 7**: [24]

$$\varepsilon_r(E_0) = \varepsilon_{rl} + \sqrt{\varepsilon_{r-rev}^2 + (\alpha_r * E_0)^2} \quad (7)$$

Where $E_o$ is the oscillating electric field, $\varepsilon_{rl}$ is the intrinsic lattice contribution, $\varepsilon_{r-rev}$ is the reversible contribution from domain wall vibrations, and $\alpha_r$ is Rayleigh´s coefficient which is a measure of the domain wall friction against the irregular energy landscape created by internal defects and disorder.[25] The value of $\alpha_r$ for the "On GSO" film is 0.57 cm.kV⁻,is higher than 0.22 cm.kV⁻ measured for the "Off GSO" membrane. Moreover, we can describe a threshold Electric field ($E_{th}$) that represents the extent of wall pinning using **Equation 8:** [24]

$$E_{th} = \frac{\varepsilon_{r-rev}}{\alpha_r} \quad (8)$$

$E_{th}$ values are calculated to be 32.3 kV.cm⁻ and 8.13 kV.cm⁻ for the film and membrane respectively, showing explicitly that the pinning field is lower in the membrane capacitor.

## 2.5. Energy Storage characteristics

The different dynamics have consequences for the energy storage behavior of the capacitors. For electric energy storage, the critical field should be close to the operating field ($E_{max}$), the remnant polarization ($P_r$) should be as low as possible and the saturation polarization ($P_s$) as high as possible. In addition, since the opening of the hysteresis loops is proportional to the energy loss, it should be minimized to maximize efficiency. These conditions can be understood by visual inspection of a hysteresis loop (**Figure 7a**), where the area shaded in green is the energy stored by the capacitor and the area shaded in pink is the energy lost. In **Figure 7b,** we compare the energy storage density and the efficiency of the on-GSO and off-GSO capacitors, calculated using **Equation 9 (a-c)** below,

$$U_{stored} = \int_0^{E_{max}} E_{return} dP \tag{9a}$$

$$U_{loss} = \int_0^{E_{max}} (E_{forward} - E_{return}) \, dP \tag{9b}$$

$$Efficiency\ (\%) = \frac{U_{stored}}{U_{stored} + U_{loss}} * 100 \tag{9c}$$

where $U_{stored}$ is the stored and recoverable energy, $U_{loss}$ is the energy loss, and $E_{max}$ is the maximum applied electric field.

As expected, both capacitor morphologies (films and membranes) show a drop of energy storage density and efficiency as the frequency of the AC loop increases, reflecting the increasing hysteresis and domain wall friction derived from the functional measurements (polarization loops and Rayleigh coefficients). Importantly, however, at all frequencies, the membrane shows higher energy storage density and efficiency than the epitaxial film thanks to its faster switching dynamics and lower losses.

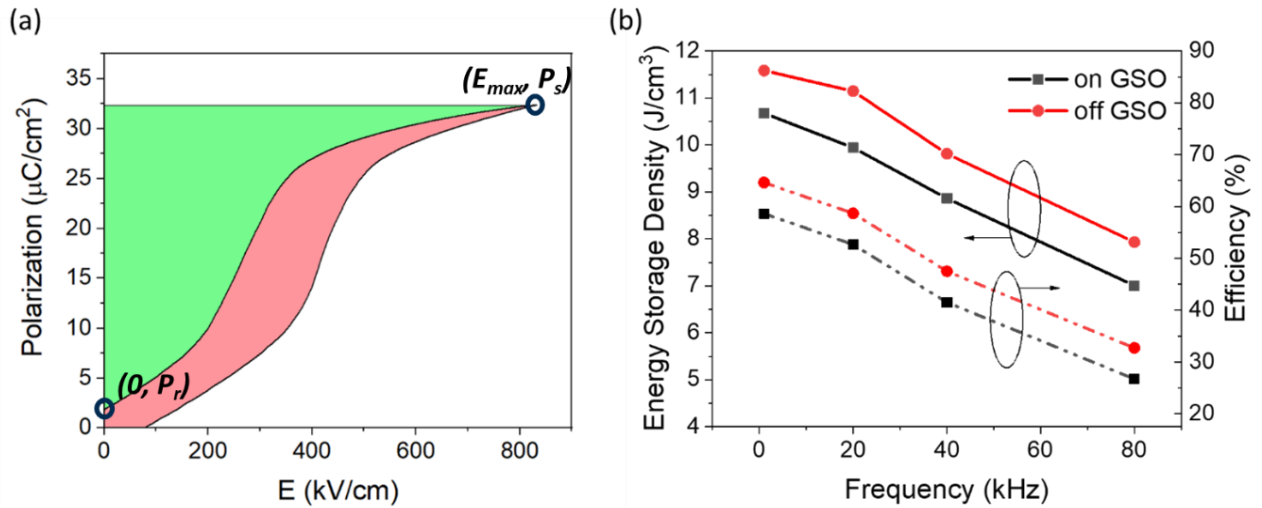

**Figure 7: (a):** Positive voltage loop (centered) of SRO/PZO/SRO membrane capacitor (off GSO) measured at 1 kHz. The green area represents the stored energy, pink area represents

the energy loss, **(b):** Energy Storage density and efficiency of epitaxial film and membrane capacitors.

## 3. Summary and Conclusion

The results demonstrate that it is possible to make free-standing antiferroelectric capacitors with better functionality than that of epitaxially-clamped devices. Although both epitaxial and free-standing capacitors show the expected Pbam structure and robust antiferroelectric hysteresis, the membrane capacitors show lower residual polarization, sharper hysteresis loops, and lower losses than the epitaxial ones. Analysis of the switching dynamics shows that the response of the membranes is more agile, with a lower frequency exponent in the KAI model and lower Rayleigh coefficient. These results are all consistent with a higher mobility/lower pinning of the AFE-FE phase boundaries in the free-standing membranes.

The sharper hysteresis and lower residual polarization result in a higher energy-storage density and efficiency for free-standing films compared to epitaxially-clamped capacitors. Combined with other morphological qualities (flexibility, transferability to Silicon supports), the results make a very good case for the device applications of antiferroelectric membranes.

## 4. Experimental Section/Methods

PZO sandwiched between SrRuO3 (SRO) heterostructures are grown either with sacrificial layer of $Sr_3Al_2O_6$ (SAO) between the heterostructure and the substrate to fabricate free-standing capacitors or without as epitaxial capacitors and $GdScO_3$_ GSO (110). These films were grown using Pulsed laser deposition (PLD) using KrF excimer laser (248 nm, COMPex 102, Lambda Physik). Bottom SRO layer was grown at 750ºC in dynamic Oxygen partial pressure of 100 mTorr with a fluence of 1.833 J $cm^{-2}$, whereas the top SRO layer was grown at same conditions except for the temperature of 575ºC, PZO was grown at 575ºC in dynamic oxygen partial pressure of 100 mTorr with a fluence of 1.833 J $cm^{-2}$, and SAO was grown at 750ºC in dynamic oxygen partial pressure of 1 mTorr with a fluence of 2.5 J $cm^{-2}$.

For the epitaxial films, the top electrodes were patterned via photolithography. The top SRO layer was spin-coated with the photoresist AZ5214E from Laurell Tech, using WS400BZ-6NPP LITE. The program for spin coating was 500 rotations per minute (rpm) for 5s and then 4000 using the software Clewin and the patterning was done using DWL Lithography using Kloé Dilase 250. For the electrodes with diameters 400 to 100 µm, the velocity used was 0.5 µm $s^-$ with a modulation of 50 and for electrodes 50 to 10 µm, 0.05 µm $s^-$ with modulation of 5. After the first UV exposure, the sample was baked at 120 ºC for 120 seconds and then again exposed completely to UV using the Kloé KLUB 3 for 10 seconds. The photoresist was then developed by AZ 726 and the SRO was etched using Sodium periodate ($NaIO_4$) chemical etchant. Finally, the sample was washed with acetone to remove all remaining photoresist.

To fabricate the membranes, the top surface of the heterostructure with SAO was attached to PDMS, and then placed in a water bath for the SAO to dissolve. After dissolution of SAO, the substrate fell off. The secondary substrate, metal coated Silicon, was heated to 40ºC and the membrane along with PDMS was brought into contact with the secondary substrate. The whole substrate was then heated to 60ºC and left for 10 minutes. Afterwards, the PDMS was lifted slowly using a micrometer probe, thereby removing the capacitor from the PDMS onto the secondary substrate.

The XRD was measured using Panalytical X'pert Pro diffractometer (Copper K-$\alpha_1$, 1.540598Å), using a hybrid 2-bounce primary monochromator on the incident beam side and a PIXcel line detector. The dielectric measurements were done using Agilent E4980A Precision LCR meter, while heating and cooling the samples in a Nextron probe station with piezo-controlled probes. A Thermo Fisher Tecnai F20 TEM was used for the low magnification imaging and Thermo

Fisher Themis-Z Double-corrected 60-300 kV S/TEM was used for high-resolution observation. The convergence and collection angle under the STEM-HAADF mode was 17.9 mrad and 50–200 mrad, respectively. The hysteresis loops and resistive-capacitance measurements were done using aixACCT TF Analyzer 2000.

**Supporting Information**
Supporting Information is available from the Wiley Online Library or from the author.

**Acknowledgements**
This project was funded by Grant PID2019-108573GB-C21 (FOx-Me) funded by the Spanish Ministry of Science and Innovation, and by grant N° 766726 (TSAR) from the European Union's Horizon 2020 research and innovation program. Ying Liu acknowledges the BIST Postdoctoral Fellowship Programme (PROBIST) funded by the European Union's Horizon 2020 research and innovation programme, under the Marie Sklodowska-Curie grant agreement No. 754510. David Pesquera acknowledges funding from 'la Caixa' Foundation fellowship (ID 100010434). The authors are grateful for the scientific and technical support from the Australian Centre for Microscopy and Microanalysis as well as the Microscopy Australia node at the University of Sydney.

Received: ((will be filled in by the editorial staff))
Revised: ((will be filled in by the editorial staff))
Published online: ((will be filled in by the editorial staff))

*Switching dynamics and improved efficiency of free-standing antiferroelectric capacitors*

Umair Saeed[1,2*], David Pesquera[1*], Ying Liu[1,3], Ignasi Fina[4], Saptam Ganguly[1], José Santiso[1], Jessica Padilla[1], José Manuel Caicedo Roque[1] , Xiaozhou Liao[3], Gustau Catalan[1,5]

[1]Catalan Institute of Nanoscience and Nanotechnology (ICN2), CSIC and BIST, Campus UAB, Bellaterra, Barcelona, 08193 Spain.
[2] Universitat Autònoma de Barcelona, Plaça Cívica, 08193 Bellaterra, Barcelona, Spain.
[3] School of Aerospace, Mechanical & Mechatronic Engineering, The University of Sydney, Sydney, NSW, 2006, Australia.
[4]Institute of Materials Science of Barcelona (ICMAB-CSIC), Campus UAB, Bellaterra, Barcelona, 08193 Spain.
[5]ICREA - Institució Catalana de Recerca i Estudis Avançats, Barcelona, Catalonia, 08010 Spain.

E-mail:    umair.saeed@icn2.cat
           gustau.catalan@icn2.cat

**S1: STEM image of ¨On GSO¨ epitaxial film.**

**(a)** Low resolution TEM image, **(b)** Lead atom displacement mapping superimposed on HAADF TEM image and inset showing zoom in of the red circle, **(c)** GPA rotation maps. The area enclosed in red circle shows a FiE region with unequal opposite Pb atom displacement, white arrow represents a translational boundary.

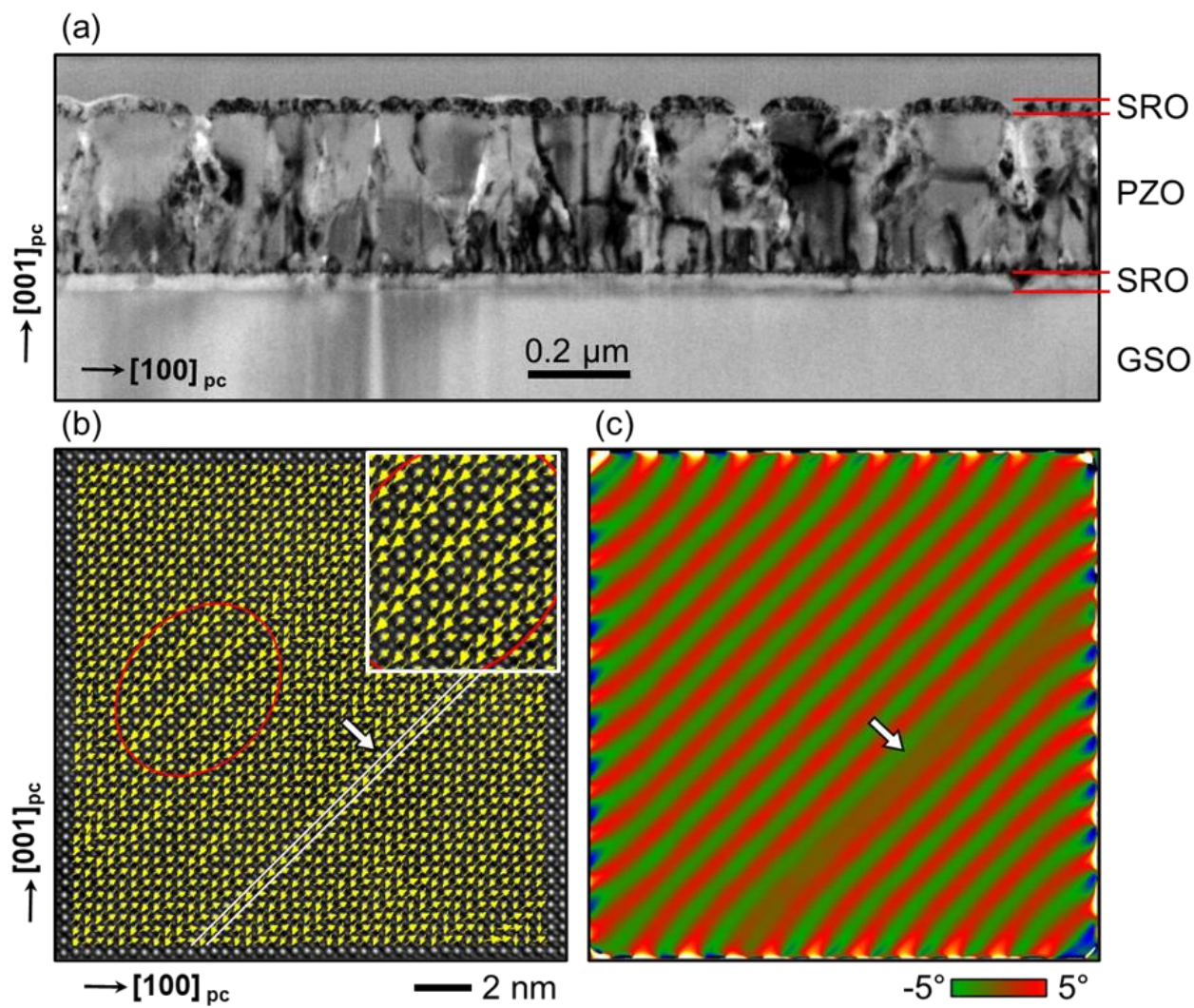

**S2: STEM of columnar boundary in "off GSO" membrane.**

The images show STEM observation of a region where the two columns intersect. We see $(100)_{pc}$ planes of one the columns are not parallel to the surface plane. The GPA of this region shows dislocations where the two columns meet, and the resulting strain in the locale. The GPA does not show a strain pattern as in **Figure 3** or **4** of the main text, as this region does not contain the right variation of the $(120)_O$ orientation. These dislocations and changes in strain state between two columns might act as pinning points for phase boundaries during electric field induced AFE to FE transition, that affects the dynamics of the transition. This might also contribute to the values of the fitting parameters of KAI model used in the main text, and lead to high values of β exponent.

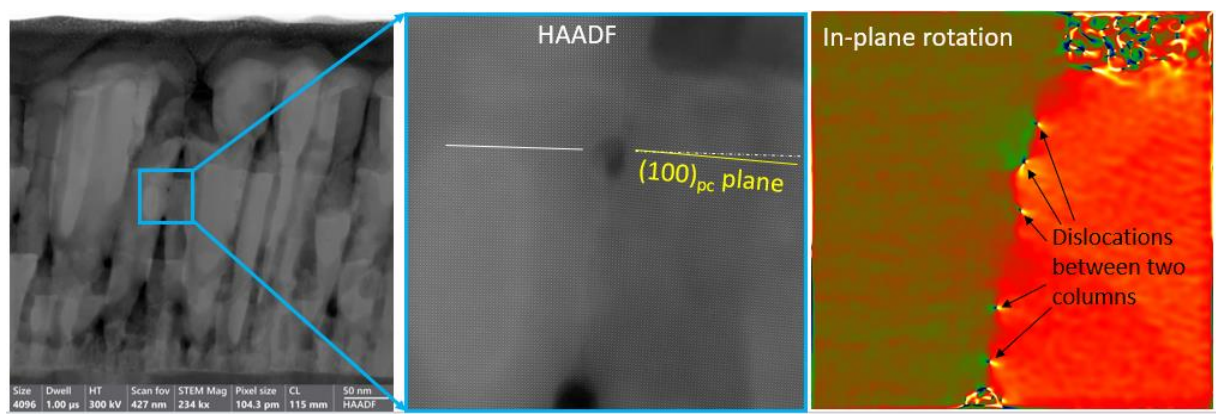

**S3: Hysteresis loops comparison of "On GSO" and "Off GSO" at different frequencies.**

The hysteresis loops shown are calculated using the integration of the current density peaks w.r.t time and plotting against Voltage.

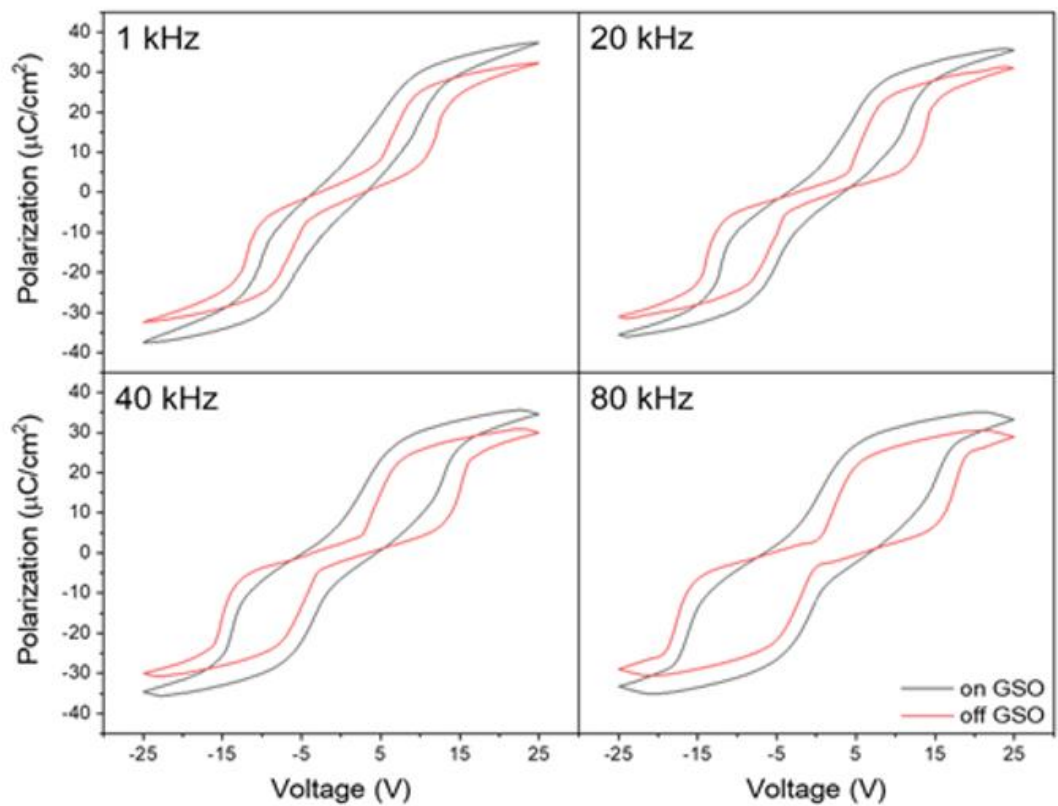

**S4: Resistive- Capacitive time constants in (a): ¨on GSO¨ and (b): ¨off GSO¨ samples.**[1]

RC effects in samples can affect the shift of the characteristic switching electric fields with change in frequency. Therefore, it is important to know the RC time constants of each sample. To measure the RC time constants, the epitaxial film "on GSO" and the membrane "off GSO" were tested by applying a 5V AC loop, at 20 kHz. 5V is chosen as the amplitude because at this voltage, no switching of the sample occurs. The frequency is chosen in a way that the preset current range is the same as the range at which the frequency dependent loops were measured at 1 mA, since this range can also affect the impedance and the change in the switching electric fields with frequency.

The drop in current with voltage is fitted with the equation (red line in the Figure):

$$I = I_0 . exp^{-\frac{t}{\tau}}$$

Tau ($\tau$) is the RC time constant and is calculated to be 453 ns in ¨on GSO¨ sample and 225 ns in ¨off GSO¨ sample. This difference in tau is due to the bottom electrodes i.e., SRO in the ¨on GSO¨ sample has higher RC effects than gold in ¨off GSO¨ samples. However, even at the highest frequency of 80 kHz, the voltage cycle has a time-period of 12 µs which is much higher than the RC time constants and hence, the shift in switching electrode fields depending on frequency can be said to have been not affected by the RC effects.

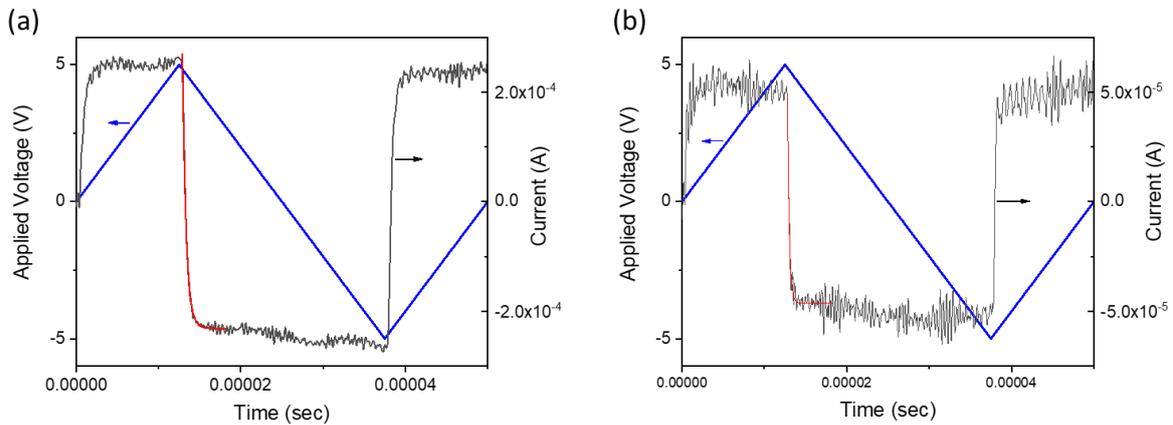

**S5: KAI model fittings for the negative electric fields of "On GSO" and "Off GSO" capacitors: (a):** AFE- FE switching field and **(b):** FE- AFE switching field.

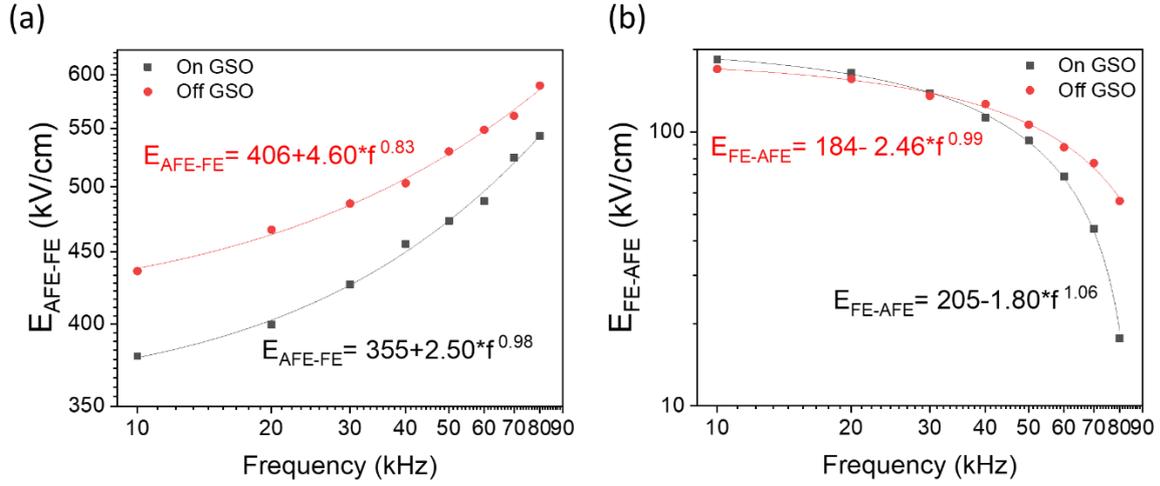

| Fitting Parameters | | On GSO | Off GSO |
|---|---|---|---|
| $E_{AFE-FE} = A + K_A * f^\beta$ | A | 355 | 406 |
| | $K_A$ | 2.50 | 4.60 |
| | β | 0.98 | 0.83 |
| $E_{FE-AFE} = F - K_F * f^\beta$ | F | 205 | 183 |
| | $K_F$ | 1.80 | 2.46 |
| | β | 1.06 | 0.99 |

**Table:** Values of fitted parameters from **equations 3 and 4**.

**S6: KAI model fittings for the smaller fitted peak in positive FE- AFE switching in ¨Off GSO¨ membrane.**
**(a):** Peak fitting on the FE- AFE current peak at 50 kHz, **(b):** KAI model fittings on the two component peaks.

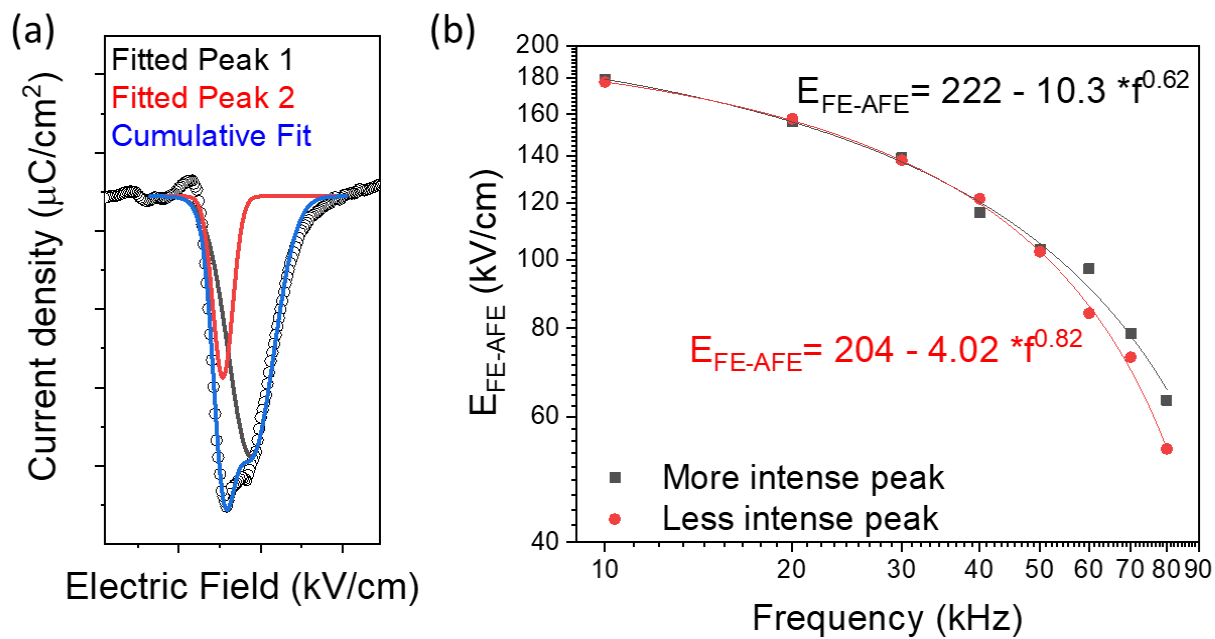

27